\documentstyle[aps,twocolumn,graphicx]{revtex}

\newcommand{\bg}[1]{\mbox{\boldmath$#1$}}

\begin{document}

\draft

\title{Dynamics of two colliding Bose-Einstein condensates\\ 
in an elongated magneto-static trap}

\author{M. Modugno$^1$, F. Dalfovo$^2$, C. Fort$^1$, P.
Maddaloni$^{1*}$, and F. Minardi$^1$} \address{
$^1$ INFM -- L.E.N.S. -- Dipartimento di Fisica, Universit\`a di Firenze\\
L.go E. Fermi 2, I-50125 Firenze, Italy\\
$^2$ Dipartimento di Matematica e Fisica, Universit\`a Cattolica,\\
and Istituto Nazionale per la Fisica della Materia, Gruppo Collegato
 di Brescia,\\ Via Musei 41, I-25121 Brescia, Italy}

\date{\today}

\maketitle

\begin{abstract}
We study the dynamics of two interacting Bose-Einstein con\-den\-sa\-tes,
by nu\-me\-ri\-cal\-ly sol\-ving two cou\-pled Gross-Pitaevskii equations
at zero temperature.  We consider the case of a sudden transfer of 
atoms between two trapped states with different magnetic moments:  
the two condensates are initially created with the same density 
profile, but are trapped into different magnetic potentials,
whose minima are vertically displaced by a distance much larger 
than the initial size of both condensates.  Then the two condensates 
begin to perform collective oscillations, undergoing a complex
evolution, characterized by collisions between the two condensates.
We investigate the effects of their mutual interaction on the 
center-of-mass oscillations and on the time evolution of the aspect ratios.  
Our theoretical analysis provides a useful insight into the recent 
experimental observations by Maddaloni {\it et al.}, cond-mat/0003402. 
\end{abstract}

\section{Introduction}

Several papers have been recently devoted to the study of 
multiple-species interacting Bose-Einstein condensates (BEC)
\cite{maddaloni,jila1,jila2,jila3,jila4,otago,williams,sinatra,sinatra2}.  
The first experiments involving the interaction between multiple 
condensates have been performed at JILA, with two condensates in the 
hyperfine levels $|F=2,m_f=2\rangle$ and $|1,-1\rangle$ of $~^{87}$Rb
confined in a Ioffe-type trap \cite{jila1}.   Other experiments have 
been performed with the states $|2,1\rangle$ and $|1,-1\rangle$ in a 
TOP trap, in a situation of nearly complete spatial overlap of the 
two condensates. In fact they have almost the same magnetic moment and 
therefore experience two (slightly displaced) trapping potentials with 
the same frequencies \cite{jila2,jila3,jila4}.  The resulting 
dynamics reveals a complex structure and it is characterized by a
strong damping of the relative motion of the two condensates
\cite{jila2}. These results have been analyzed in \cite{sinatra}, by 
identifying two  dynamical regimes: (i) a periodic motion with a slow 
frequency (with  respect to the trap frequency), for small displacement 
between the  traps; (ii) a strong non-linear mixing which leads to a 
damping of the  relative motion, for larger displacements (but still 
smaller than the  size of the initial condensate).  By comparing  
the numerical solution of the Gross-Pitaevskii equation with the 
JILA experiment \cite{jila2}, the authors of Ref.~\cite{sinatra} 
find a fair agreement, although small oscillations  remain undamped 
for longer times compared to the experimental  observation 
\cite{sinatra,sinatra2}. 

A mixture of $^{87}$Rb condensates in different $m_f$ states in a TOP 
trap has been experimentally investigated also by another group 
\cite{otago}, but no effects of the mutual interaction have been 
observed. 

Recently we have reported some experiments performed at LENS with two 
interacting condensates \cite{maddaloni}, which reveal several new
features with respect to those analyzed in Refs.~\cite{jila2,sinatra}. 
In fact, we have considered a sudden transfer of atoms between the 
states $|2,2\rangle$ and $|2,1\rangle$ of $~^{87}$Rb, having different
magnetic moments. The $|2,1\rangle$, which is created with the same 
density distribution and at the same  position of the $|2,2\rangle$
condensate, feels a magnetic potential whose minimum is displaced
along the vertical $y$ axis by  a distance much larger that the initial
size of the condensate. Therefore, its dynamics is characterized by large
center-of-mass oscillations. The  two condensates periodically collide and
these collisions have significant observable effects. Among them:  i) the
shape oscillations of $|2,2\rangle$ condensate, triggered by the sudden
transfer to the $|2,1\rangle$ state, are significantly enhanced; ii) the
center-of-mass oscillation frequency of the $|2,1\rangle$ condensate is
shifted upwards.  

By means of preliminary numerical simulations, we have 
already shown that the basic features of these experimental 
observations can be explained within the Gross-Pitaevskii (GP) 
theory \cite{maddaloni}. In this paper we better analyse the 
theoretical scheme and describe in more detail the numerical 
procedures and the approximations made.

The paper is organized as follows:  in Section \ref{sec:theor}, we
introduce the two coupled GP equations for the dynamics of the 
condensates and discuss the approximations used to solve them.  In
Section \ref{sec:cm} we show the results for the center-of-mass
motion, stressing the effect of mutual interactions in 
the oscillation frequency. In Section \ref{sec:aspect} we discuss
the collective modes of each condensate and the time evolution of
the aspect ratios. In Section \ref{sec:expansion} we simulate the
expansion of the two condensates after switching-off the trapping
potential. In this way, we can compare our theoretical results
with the experimental data \cite{maddaloni} for the center-of-mass
motion and the aspect ratios.

\section{Gross-Pitaevskii theory for two coupled condensates}
\label{sec:theor}

At very low temperatures, when there is no thermal cloud,
the evolution of the two interacting Bose-Einstein condensates 
can be described by two coupled Gross-Pitaevskii (GP) equations 
($i=1,2$):
\begin{equation}
i\hbar{\partial\Psi_i\over\partial t} = \left[-
{\hbar^2\nabla^2\over2m} + V_i
+\sum_{j=1,2} {4\pi\hbar^2 a_{ij}\over m}|\Psi_j|^2 \right]
\Psi_i \; . 
\label{eq:gp0}
\end{equation}
Here, $\Psi_1$ and $\Psi_2$ are the order parameters of the
$|2,1\rangle$ and $|2,2\rangle$ condensates of $~^{87}$Rb 
(hereafter called $|1\rangle$ and $|2\rangle$). By solving 
Eq.~(\ref{eq:gp0}) we want to simulate the situation in 
which a certain amount of atoms are suddenly transferred from 
the condensate $|2\rangle$, initially in its ground state in 
the trap, to the condensate $|1\rangle$. Due to the different 
magnetic moment, the condensate $|1\rangle$ experiences a trapping 
potential having the axial and radial frequencies lowered by a 
factor $\sqrt{2}$ and the minimum shifted radially by a constant 
value. Thus, the trapping potential $V_i$ are different and can be
written as
\begin{eqnarray}
V_1(x,y,z) &=& {m\over2}\omega_{\perp1}^2 \left[x^2 + (y+y_0)^2
+ \lambda^2 z^2\right]\\
V_2(x,y,z) &=& {m\over2}\omega_{\perp2}^2 \left[x^2 + y^2 
+ \lambda^2 z^2\right] \; . 
\end{eqnarray}
 From now on  
we use the same parameters as in the experiments of  
Ref.~\cite{maddaloni}, that is: scattering lengths $a_{22}=a_{12}
\simeq 98 a_0$ and $a_{11}\simeq 94.8 a_0$, trapping 
frequencies $\omega_{\perp2}=2\pi \times 164.5$ Hz,  
$\omega_{\perp1} = \omega_{\perp2}/\sqrt{2} = 2 \pi \times 116.3$ 
Hz, and displacement $y_0 = g/ \omega_{\perp2}^2= 9.2 \mu$m.  
The two condensates have the same asymmetry parameter 
$\lambda \equiv \omega_{z}/ \omega_{\perp}\simeq~0.0766$.  
For the total number of atoms we use $N=N_1+N_2=1.5 \times 10^{5}$ 
with $N_1=0.13N$. 

In order to numerically solve the GP equations (\ref{eq:gp0}) one 
should map the two order parameters $\Psi_i$ on a fully 
three-dimensional grid of points. This should be large enough to 
describe the oscillations of each condensate in the trap 
(center-of-mass motion plus internal excitations) as well as their 
expansion when the trap is switched off. This approach is 
computationally heavy. We prefer, at this stage, to solve the 
GP equations (\ref{eq:gp0}) within different approximate 
schemes, depending on the type of process we want to describe.
We distinguish the following three cases.

Case A:  dynamics of a condensate in the trap ignoring its 
interaction with the other. In this case, the center-of-mass (CM) 
motion is trivial: the CM of the condensate $|2\rangle$ remains at 
rest, while the one of  $|1\rangle$ oscillates freely around $y_0$
at the frequency $\omega_{\perp1}$. The internal motion of each
condensate originates from the fact that its starting configuration,
after transferring atoms from $|2\rangle$ to $|1\rangle$, is 
not the ground state configuration. In particular, the state $|2\rangle$ 
feels the same trap but has less atoms: its starting shape is wider than 
the equilibrium one. Its internal motion  can be predicted 
by exactly solving the corresponding GP equation, without coupling 
with $|1\rangle$, i.e., with $a_{12}=0$. In this case, the equation 
has axial symmetry and we can use the numerical code developed by 
two of us in Ref.\cite{modugno}. 
 
Case B: effects of the mutual interaction in the radial motion. 
In order to point out the basic effects of the interaction occurring 
between the two condensates when they collide, one can decouple the
slow axial motion from the GP equation. In this way, the simulation
is reduced to a two-dimensional problem in the $xy$-plane, which is
easier to perform. We solve the coupled GP equations by using the 
Split-Step Fourier method \cite{jackson}, with a Fast Fourier 
Transform algorithm \cite{recipes}, by using cylindrical coordinates
and dimensionless variables, as explained in the Appendix. We will 
justify the decoupling of the axial motion later, when discussing 
the low energy collective modes of these condensate. 

Case C: dynamics of the free expansion. In the experiments, the
CM positions and the aspect ratios of the condensates are measured
after few milliseconds of expansion, with the trap switched off. 
We simulate the expansion, starting from the configuration obtained
in case A or B at a given time. The CM of the two condensates simply 
undergo a free fall under gravity. The shape of each condensate  
changes during the expansion. We simulate the expansion of the condensate 
$|2\rangle$ by using a scaling law valid in the Thomas-Fermi
(large $N$) limit \cite{castin,bec_review}. We instead ignore 
the mutual interaction of the condensates in this process; 
even if they overlap after some time (see Section \ref{sec:expansion}), 
their density is much lower than the initial one, due to the fast radial 
expansion, and hence the mean-field interaction is also much
weaker. 

\section{Center-of-mass motion in the trap}
\label{sec:cm}

We start from an initial configuration corresponding to the 
lowest stationary solution 
of Eq.\ (\ref{eq:gp0}), with $N$ atoms in the $|2\rangle$ condensate. 
This can be obtained by iterating in imaginary time, starting from a
trial wave function, as done in \cite{dalfovo}. Then, at an arbitrary 
time $t=0$, one builds two condensates $|1\rangle$ and $|2\rangle$, with 
the same density profile of the starting one, but with different number 
of atoms, $N_1$ and $N_2$.  

If they do not interact ($a_{12}=0$), 
the CM of condensate $|2\rangle$ remains at rest, while the one of the
condensate $|1\rangle$ starts oscillating around its equilibrium position
$y=-y_0$ with frequency $\omega_{\perp1}$, as shown in Fig.~\ref{fig:cm-trapped}a.

In order to include the effects of the mutual interaction, we 
make a two-dimensional simulation as explained in case B of the 
previous Section, by ignoring the axial motion and solving the two 
coupled GP equations in the $xy$-plane. Thus, the initial condensate 
is an infinite cylinder along $z$ while the normalization of the 
two-dimensional order parameter in the $xy$-plane is chosen in such 
a way to preserve the profile in the radial direction, when passing 
from $3$ to $2$ dimensions, as explained in the Appendix. The results
are shown in Fig.~\ref{fig:cm-trapped}b. 
One sees that now  both condensates move. The 
motion of the condensate $|2\rangle$ originates from the mutual 
repulsion with the condensate $|1\rangle$, which acts both at the
beginning, when $|1\rangle$ is formed and starts moving away from 
$|2\rangle$, and when they periodically collide. The 
motion of $|1\rangle$ is still dominated by large oscillations
around $-y_0$, but their frequency $\omega_{1}$ turns out to be 
larger than the non-interacting frequency $\omega_{\perp1}$ by about $5.4\%$. 
Furthermore, the oscillations appear to be damped. 
We notice that, though the minimum separation between the two 
center-of-mass positions slowly grows in time, it remains always 
smaller than  the sum of the two condensate radii, at least in time 
interval here considered,  thus ensuring that the two condensates 
indeed collide.

The shift in frequency of the CM oscillations is one of the clear
signatures of the the mutual interaction. It originates from the
fact that the condensate $|1\rangle$ moves in an effective trap
which is anharmonic. In fact, the presence of the condensate
$|2\rangle$ at one side of the harmonic potential in which 
$|1\rangle$ moves, provide an extra repulsion with a consequent
up-shift of frequency. This shift only occurs if the two 
condensates overlap periodically. 
Since it comes from the cross mean-field term in the 
GP equation, it depends significantly on $N_2$.  
 
Indeed, as a first approach we can consider the 1D problem of one 
single classical particle moving in the effective potential $V_e(y)$ given 
by the harmonic oscillator plus the repulsion of the $|2\rangle$ 
atoms: 
\begin{equation} 
V_e(y) = {m\over2}\omega_{\perp1}^2 y^2 
+ g_{21} n_0 \max\left\{0,1-{(y+y_0)^2\over R_{\perp2}^2}\right\} 
\end{equation} 
where $g_{21}=4\pi\hbar^2 a_{21}/m$, while 
$R_{\perp2}$ and $n_0$ are the 
radius and the central density of the $|2\rangle$ condensate 
in the Thomas-Fermi (large $N$) limit, respectively \cite{bec_review}. 
To obtain the oscillation period, we carry out two integrals 
\begin{eqnarray} 
T &=& \sqrt{2m}\int_{-y_0}^{-y_0+R_{\perp2}}\hspace{-0.5cm}dy 
\left[V_e(-y_0)- {m\over2}\omega_{\perp1}^2 y^2\right.
\nonumber \\ 
&&\left.\qquad \qquad +g_{21} n_0 
\left(1-{(y+y_0)^2\over R_{\perp2}^2}\right)\right]^{-1/2}
\nonumber \\  
&&+\sqrt{2m}\int_{-y_0+R_{\perp2}} ^{y_i}\hspace{-0.5cm}dy 
\left(V_e(-y_0)- {m\over2} \omega_{\perp1}^2 y^2 \right)^{-1/2} 
\end{eqnarray} 
where $y_i$ indicates the classical inversion point, equal to 
$\sqrt{y_0^2+2R_{\perp2}^2}$. 
We find 
\begin{eqnarray} 
T&=&{2\over \omega_{\perp1}} \left[{\pi\over2}+\arcsin\left(
{1-R_{\perp2}/y_0 \over 
  \sqrt{1+2R_{\perp2}^2/y_0^2}}\right) \right.\nonumber \\ 
&&\left.\qquad \qquad+ {\rm arcosh} 
\left(1+{R_{\perp2}\over y_0}\right) \right]. 
\label{e:classicT} 
\end{eqnarray} 
The period depends on the number of atoms only through the radius 
$R_{\perp2}$ that scales as $N_2^{1/5}$. In Fig. \ref{fig:classic} 
we see that the frequency shift predicted by Eq.~(\ref{e:classicT}) is
nicely close to the GP results. 

One can better visualize the effects of the interactions in the
collisions by plotting the various energy terms contributing to 
the total energy of the system, as in Fig.~\ref{fig:energy}.
We checked that the 
total energy is accurately conserved during the simulation. The
energies of each condensate (kinetic, mean-field, trapping) vary
with time in a characteristic way: they change only when the 
condensates collide. This happens about every $8$~ms.  
During the short collision time, when they 
partially overlap, the mutual interaction energy, i.e., the term 
containing the scattering length $a_{12}$, is significantly 
non-zero. In the upper part of the same figure we redraw the
CM position of the condensate $|2\rangle$, with a vertical scale 
larger than in Fig.~\ref{fig:cm-trapped}. 
On notices that the slope of these  
oscillations changes whenever the two condensates collide, the
precise change (sign and amplitude) depending on the phase of the 
oscillation when each collision occurs.     

The behavior of the energy {\it vs.} time is strictly related to 
the damping of the CM oscillations of condensate $|1\rangle$, shown 
in Fig.~\ref{fig:cm-trapped}b. 
First, we remind that in the GP theory the damping is not 
due to dissipative processes, since the total energy is conserved. 
What happens, instead, is that the kinetic energy initially associated 
with the CM motion is eventually shared among many degrees of freedom. 
This redistribution of energy is caused by the periodic collisions. 
One of these degrees of freedom is the CM motion of the condensate 
$|2\rangle$. But the internal excited states of each condensate are 
also expected to play an important role, as discussed in the next 
section.

\section{Condensate deformations and aspect ratios in the trap}
\label{sec:aspect}

Let us consider the internal motion of the condensate $|2\rangle$.
Just after the transfer of $N_1$ atoms into $|1\rangle$, its density
distribution has the same profile as the ground state of $N$ atoms,
but with $13$\% of atoms missing. Thus, it starts oscillating around 
the new equilibrium configuration, the confining potential being 
unchanged. This situation is very similar to that studied in 
Ref.~\cite{jila5}, where similar oscillations were produced by a
sudden change of scattering length. 

If one ignores the interaction with the condensate $|1\rangle$, the
motion of the condensate $|2\rangle$ can be predicted by solving the 
GP equation (\ref{eq:gp0}) with $i=2$ and $a_{12}=0$. This corresponds
to case A in Section \ref{sec:theor}. In Fig.~\ref{fig:aspect-nonint}, 
we show the resulting aspect ratio, $[\langle y^2 \rangle / \langle z^2 
\rangle ]^{1/2}$ (solid line). The curve is clearly a superpositions
of two, fast and slow, modes. 

To understand which excited states partecipate in the curve  
in Fig.~\ref{fig:aspect-nonint}, we notices that, since the new 
equilibrium state is not far from the starting configuration, the 
amplitude of the induced oscillations is relatively small. Therefore, 
only the lowest energy modes are excited and, among them, the 
ones with no angular momentum. For an elliptic trap, where only the 
$z$-component of the angular  momentum is conserved, two $m=0$ 
states are expected to be excited. The one with lowest frequency 
corresponds to in-phase oscillations of the width along $x$ and 
$y$ and out-of-phase along $z$. The one with highest frequency is 
an in-phase compressional mode along all directions (breathing mode). 
In the Thomas-Fermi (TF), large $N$, regime and in the linear limit
(small amplitude), their frequency is 
\cite{stringari}
\begin{equation}
\omega_{\mp} = \sqrt{q_{\mp}}\ \omega_{\perp2} 
\label{eq:omega-mp}
\end{equation}
with 
\begin{equation}
q_{\mp} = 2+ (3/2)  \lambda^{2} \mp (1/2) 
\sqrt{ 9 \lambda^{4}-16 \lambda^{2}+16 } \; . 
\label{eq:qmp}
\end{equation}
When the condensate is also strongly elongated ($\lambda \ll 1$) the
two frequencies become
\begin{equation}
\omega_{+}\simeq2 ~\omega_{\perp2} \; ; \qquad 
\omega_{-}\simeq \sqrt{5\over2} ~\lambda~\omega_{\perp2} \; .
\label{eq:lambda-0}
\end{equation}
In this limit the two frequencies are quite different, and the axial 
and radial collective excitations are almost decoupled. The radial 
width is characterized by small oscillations of frequency $\omega_{-}$ 
superimposed to a wider oscillation of frequency $\omega_{+}$, 
and vice-versa for the axial width \cite{ketterle}. Using the
parameters for the condensate $|2\rangle$ in 
Eqs.~(\ref{eq:omega-mp})-(\ref{eq:qmp}), one finds values
which differ less than $0.04 \%$
from their asymptotic $\lambda \to 0$ values in Eq. (\ref{eq:lambda-0}). 

The TF approximation corresponds to neglecting the laplacian 
of the density (i.e., the quantum pressure) in the GP equation. 
In this case, the density profile of the condensate in each direction 
is an inverted parabola which vanishes at the classical turning point. 
In the radial and axial directions, these points, 
$R_\perp$ and $Z$, are given by the condition $\mu=m\omega_{\perp2}^2 
R_{\perp}^2/2=m\omega_{z2}^2 Z^2/2$, where $\mu$ is the chemical 
potential fixed by the normalization of the order parameter to $N_2$. 
A nice feature of the TF regime is that both the free
expansion and the lowest collective modes exhibit an exact scaling
behavior: the parabolic shape is preserved and the widths $R_\perp$ 
and $Z$ scale as  $R_{\perp}(t)=R_{\perp}(0) b_\perp(t)$ and 
$Z(t)=Z(0) b_z(t)$. The GP equation for the condensate $|2\rangle$
then transforms into two coupled equations for the scaling parameters
$b_\perp(t)$ and $b_z(t)$: 
\begin{eqnarray}
\ddot{b}_\perp & + & \omega_\perp^2 b_\perp - 
{ \omega_{\perp}^2 \over b_\perp^3 b_z } = 0 \\
\ddot{b}_z & + & \omega_z^2 b_z - 
{ \omega_{z}^2 \over b_\perp^2b_z^2 } = 0 \; . 
\label{eq:ddotb}
\end{eqnarray}
The numerical solution of these differential equations is 
straitghforward. The result for the evolution of the aspect ratio 
of condensate $|2\rangle$, $R_{\perp}/Z$, is shown as a dashed line 
in Fig.~\ref{fig:aspect-nonint} and compared with the one obtained 
with the exact solution  of the GP equation (solid line) for the
same condensate, $[\langle y^2 \rangle / \langle z^2  \rangle ]^{1/2}$. 
The shape of the two curves is very similar as a consequence of the 
fact that the parameter $N a / a_{\rm ho}$, with $a_{ho} = [ \hbar 
/ (m\omega_{ho} ]^{1/2}$, is indeed much greater than $1$, as required 
to ensure the validity of the TF approximation. They differ only 
for a small vertical shift due to the different definitions
of the aspect ratio; this shift would vanish only in the $N \to 
\infty$ limit \cite{note}. 

The aspect ratio in Fig.~\ref{fig:aspect-nonint} ignores the
mutual interaction between the two condensates in the periodic
collisions. Since this interaction breaks the axial symmetry in
the motion of $|2\rangle$, one should solve the coupled GP 
equation in full tree-dimensions. Here we proceed differently. 
We observe that the axial motion is much slower than both
the time interval between two collisions and the duration of
a single collision. We thus assume that the axial motion does
not respond significantly to the mutual interaction in the
time scale here considered. Vice-versa, the collision are 
expected to significantly affect the radial motion. We treat 
the radial motion by solving numerically the GP equation in the
$xy$-plane, including the mutual interaction (case B of Section
\ref{sec:theor}). In this case, the collisions with the
$|1\rangle$ condensate make the radial motion of $|2\rangle$
rather complex. Assuming the axial width to be the same as
without collisions, and using the radial width 
obtained by the 2D integration of the coupled GP equations, 
one can calculate the new aspect ratio. This is given in 
Fig.~\ref{fig:aspect-int}. One notices that, similarly to what 
we have found for the center-of-mass motion in Fig.~\ref{fig:cm-trapped}b,
the shape of the oscillation change at each collision, every 
$8$~ms. The evolution of the aspect ratio has a complex
structure and it is not a simple superposition of the two
oscillations  $\omega_{-}$ and $\omega_{+}$ of the non-interacting 
case.  The overall amplitude is enhanced. This reflects the 
fact that the energy of the CM oscillation of $|1\rangle$ is 
transferred, not only to the CM motion of $|2\rangle$, but also 
to the internal degrees of freedom. This again, may explain part of 
the damping of the CM oscillations of $|1\rangle$ in 
Fig.~\ref{fig:cm-trapped}b. 

Looking at Fig.~\ref{fig:aspect-int}, one sees also that 
the interaction between the two condensates produces an initial
delay for the onset of collective exitations, compared to the 
non-interacting case. This originates from the fact that the 
two condensates are initially created with the same density profile, 
and the scattering lenghts $a_{22}$ and $a_{12}$ have the same value.
Therefore the $|2\rangle$ condensate ``realizes'' to be 
out-of-equilibrium only when the other condensate has moved away,
i.e. after a time of the order of $1 \div 2$ ms. The same effect can 
be seen in Fig.~\ref{fig:energy} for the center-of-mass motion.

\section{Free expansion and comparison with experiments}
\label{sec:expansion}

In the experiments of Ref.~\cite{maddaloni} the center-of-mass 
motion and the aspect ratios are measured after releasing the 
trap and letting the condensate expand under the effect of
gravity only. 

In order to compare our theoretical predictions  with the 
experimental data we have to include the effects of the
expansion. For the center-of-mass motion this is trivial. One has
only to take care of the fact that the switch-off of the trapping
potential is not istantaneous and can produce a magnetic gradient
which affects the initial velocity of the two condensates.  The 
two velocities have been measured experimentally, giving 
$v_{1y}=0.7\pm0.1~$cm/s and $v_{2y}=1.4\pm0.1~$cm/s. Due to
this difference, the two condensates fall at different speed and
they eventually overlap for some time during the expansion. As
already said in Section \ref{sec:theor} (case C), we ignore the
mutual mean-field interaction during this process. If the
overlap does not occur too early, the condensates have enough 
time to expand fast in the radial direction, the one of the 
stronger initial confinement, and their density drops quickly, 
so that the cross mean-field energy becomes negligible. 

Our results for the CM position of the two condensates are
shown in Fig.~\ref{fig:cm}. In the upper part (a), we compare
theory and experiment in a simple case, in which all the
atoms are initially transferred into the condensate $|1\rangle$,
which then oscillates alone in its new trap. The oscillation is
undamped and its frequency is simply the trapping frequency 
$\omega_{\perp1} = 2 \pi \times 116.3$ Hz. 
This picture, in which the mean-field 
interactions does not enter at all, has been used to fix the
small delay caused by the non istantaneous switch-off of the 
trapping potential. This has been done by introducing the
delay as a fitting parameter in the theoretical curve. Its
value is $\simeq 0.6$~ms.

In the lower part of Figs.~\ref{fig:cm}, we compare theory 
and experiment for the case of the two colliding condensates.
The CM dynamics of the two condensates in the trap has been
discussed in Section \ref{sec:cm}. Starting from each 
point in Fig.~\ref{fig:cm-trapped}b, we have then added the effect
of the fall in the gravity field. At the time of the observation
(after $29.3$ ms), the condensate $|1\rangle$ is almost 
always above the condensate $|2\rangle$, which means that 
they crossed during the expansion.  

There is an overall qualitative agreement between theory and
experiments. The CM of the condensate $|2\rangle$ moves, due
to the interaction with the other condensate. The amplitude of 
the oscillation is in agreement with theory, even though more 
precise measurements are needed to make the comparison 
more quantitative. The CM of the condensate $|1\rangle$ 
exhibits a damped oscillation. Its frequency is up-shifted 
compared to the case with no collisions (upper part of the 
same figure).  The measured frequency in the case of colliding 
condensates is $123.9(3)$~Hz, with a shift about $6.4(3)$\%, which
is close to the theoretical shift, $5.4 $\%. Furthermore, the 
oscillations appear now damped with an exponential decay time of 
about $60$~ms. Our theoretical prediction for this decay time
is about $170$~ms. As already said in the previous sections, the
predicted damping originates from inelastic processes in 
which the energy of the CM motion is transferred to other 
degrees of freedom. The fact that our theory underestimate the
decay time is reasonable if one consider that the theoretical 
curve in Fig.~\ref{fig:cm} only includes the radial degrees of 
freedom (we solve the coupled GP equations in the $xy$-plane).
The remaining contribution to the damping can naturally arise
from the excitations of axial modes, which are ignored in the
present analysis, namely, the possible distortions of the shape
of the two condensates along $z$, during each collision. A source
of damping can also be the occurrence of elastic scattering 
processes between atoms of the two overlapping condensates, 
as the one discussed in \cite{nist}. These processes are not
included in the GP mean-field theory.   

In order to compare theory and experiments for the measured 
aspect ratio, one has to include the effect of the mean-field 
interaction in the expansion. For the condensate $|2\rangle$ 
one can safely use the TF approximation. In practice,
one has to solve the coupled equations (\ref{eq:ddotb}) for the 
scaling parameters $b_\perp$ and $b_z$, after dropping the term
linear in $b$, which represents the trapping potential. The 
initial conditions for ${b}(0)$ and $\dot{b}(0)$ are fixed 
by the configurations of the condensate $|2\rangle$ in the
trap at a given time $t$, as calculated in Section \ref{sec:aspect}
(see Fig.~\ref{fig:aspect-int}).   In Fig. \ref{fig:aspect2} we plot 
our results, as a function of $t$, compared with the experimental data. 
The oscillations are rather complex, as a result of the periodic 
collisions between the condensate. The average  amplitude is 
in agreement with the observed data. The slow  axial modulation 
is visible and the faster radial motion has sudden changes of 
amplitude. Considering that in this comparison there are no 
fitting parameters, and the theory contains some important 
approximations, the overall agreement is good, and provides a 
further evidence of the mutual interaction between the two 
colliding condensates.  

Finally, let us comment on the aspect ratio of the condensate 
$|1\rangle$.  This quantity is expected to follow an even more 
complex evolution, due to the fact that the initial configuration 
is very far from equilibrium: the trap frequencies are changed by 
a $\sqrt{2}$ factor, the condensate is initially displaced from 
the minimum of its trap and has a shape originally build for $N$ 
atoms instead of $N_1$. In the TF limit these reduction of the 
trapping frequencies and of the number of atoms would be equivalent 
to  an ``effective" sudden change of the trapping frequencies, of 
about $96$\%, if $N$ were kept fixed. Therefore, due to the 
non-linearity of the GP equation, the resulting evolution
might be chaotic, as predicted in Ref. \cite{kagan}. Even though
the TF approximation is rather poor for this condensate, which 
has about $20000$ atoms, we solved the TF equations for the 
evolution of $R_{\perp}/Z$ finding rather irregular oscillations
between the extreme values $1\div 15$. The comparison with experimental 
data is thus difficult, even because aspect ratios greater than about 
$5$ are not easy to be measured with the finite resolution of 
the imaging devices of Ref.~\cite{maddaloni}. In view of the
characterization of the nonlinear (possibly chaotic) dynamics 
of these condensates, we think that a deeper quantitative 
analysis of this problem is worth doing in the next future by 
solving the GP equation in full 3D.

\acknowledgments 

The authors aknowledge useful discussions with M. Inguscio and S.Stringari. 
F.D. would like to thank the Dipartimento di Fisica dell'Universit\`a 
di Trento for the hospitaly.

\appendix
\section{solving the time-dependent GP equation}
\label{sec:numerics}

The GP equations in Eq.\ (\ref{eq:gp0}) can be conveniently rewritten
in terms of dimensionless quantities, by rescaling all time and space
variables by $\omega_{\perp}\equiv\omega_{\perp2}$ and
$a_{\perp}\equiv\sqrt{\hbar/m\omega_{\perp}}$ respectively.  In fact,
by defining
\begin{equation}
\xi_1\equiv x/a_{\perp};\qquad \xi_2\equiv y/a_{\perp};\qquad 
\xi_3\equiv z/a_{\perp}
\end{equation}
\begin{equation}
\tau\equiv \omega_{\perp} t/2;\qquad 
\psi_i\equiv\sqrt{a_{\perp}^3\over N}\Psi_i;\qquad 
u_{ij}\equiv 8\pi N_j a_{ij}/a_{\perp}
\label{eq:param2}
\end{equation}
 we can write Eq.\ (\ref{eq:gp0}) as follows
\begin{equation}
i{\partial\psi_i({\bg \xi},\tau)\over\partial \tau} = \left[-\nabla^2_\xi
+V_i +\sum_{j=1,2}u_{ij}|\psi_j|^2 \right]\psi_i({\bg \xi},\tau)
\label{eq:gptime}
\end{equation}
where the magnetic trapping potential are given by ($\lambda\equiv
\omega_{z2}/\omega_{\perp2}$)
\begin{eqnarray}
V_1({\bg \xi}) &=& {1\over2}\left[ \xi_1^2 + (\xi_2+\xi_{20})^2
+ \lambda^2 \xi_3^2\right] \\
V_2({\bg \xi}) &=& \xi_1^2 + \xi_2^2
+ \lambda^2 \xi_3^2,
\end{eqnarray}
and the wave functions $\psi_i$ satisfy the normalization condition ($i=1,2$)
\begin{equation}
\int d^3\xi \ |\psi_i (\xi) |^2 = 1.
\end{equation}

One of the scheme we use (see Section \ref{sec:theor}) consists in 
reducing the dynamics to the $xy$-plane, by neglecting the axial
motion. This corresponds to solve the dynamics of the 
trapped oscillations in the limit $\lambda=0$, i.e., assuming 
a uniform wave function along the axial direction.
We write $\psi\equiv A\phi(\xi_1,\xi_2)$, and we fix $A$ 
in order to preserve the width of the condensate profile in the 
radial direction, when passing from $3$ to $2$ dimensions (in the TF limit):
\begin{equation}
A ={(15\lambda)^{2/5}\over4}\left({ N a_{22}\over a_{\perp}}\right)^ {-1/10}
\; .
\end{equation}
Then the GP equations (\ref{eq:gptime}) become
\begin{equation}
\cases{\displaystyle
i{\partial\phi_1\over\partial \tau} = \left[-\nabla^2
+{1\over2}\left(\xi_1^2 + \xi_2^2\right) 
+A^2 \sum_{j=1,2}u_{1j}|\phi_j|^2 \right]\phi_1\cr
~\cr\displaystyle
i{\partial\phi_2\over\partial \tau} = \left[-\nabla^2
+\xi_1^2 + (\xi_2-\xi_{20})^2 +A^2 \sum_{j=1,2}u_{2j}
|\phi_j|^2 \right]\phi_2}
\label{eq:gp2d}
\end{equation}
and we solve them by using the Split-Step Fourier method \cite{jackson},
with a Fast Fourier Transform algorithm \cite{recipes}, and
with the normalization condition ($i=1,2$)
\begin{equation}
\int d^2\xi\ |\phi_i(\xi)|^2 = 1 \; .
\end{equation}



\begin{figure}
\centerline{\includegraphics[width=8cm,clip=]{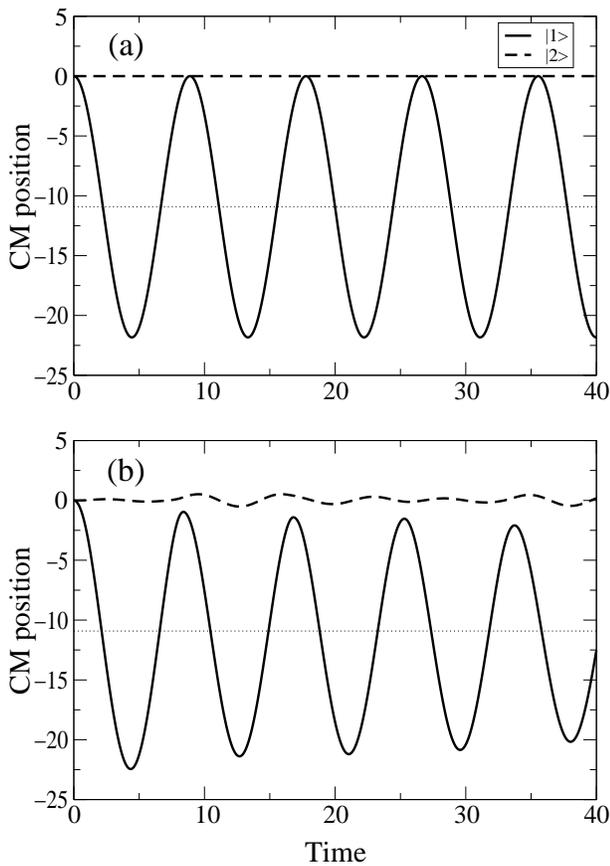}}
\caption{Center-of-mass oscillations of the $|1\rangle$ (solid line) 
and $|2\rangle$ (dashed line) condensates in the trap: 
(a) ignoring the mutual interaction, (b) including the mutual interaction. 
The dotted line corresponds to the equilibrium position $y=-y_0$ of 
the $|1\rangle$ condensate. Time and 
lengths are given in units of $\omega_{\perp2}^{-1}$ and 
$a_{\perp2}= [\hbar/(m\omega_{\perp2})]^{1/2}$, respectively.
}
\label{fig:cm-trapped}
\end{figure}

\begin{figure}
\centerline{\includegraphics[width=8cm,clip=]{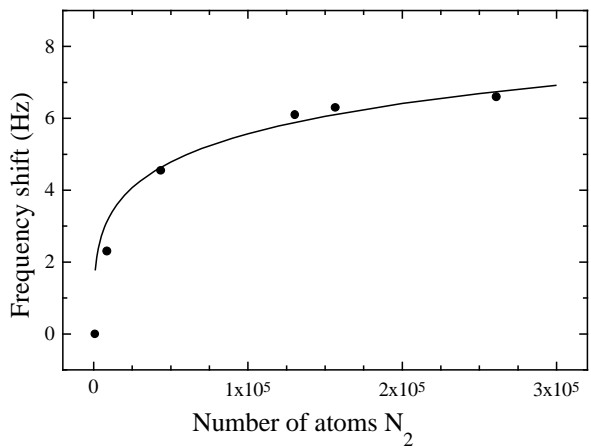}}
\caption{Frequency shift of the center-of-mass oscillation of  the
$|1\rangle$ condensate as a function of the number of atoms $N_2$ remaining in 
$|2\rangle$. GP results (points) are compared with the
classical prediction of Eq. (\ref{e:classicT}) (solid line).}
\label{fig:classic}
\end{figure}

\begin{figure}
\centerline{\includegraphics[width=8cm,clip=]{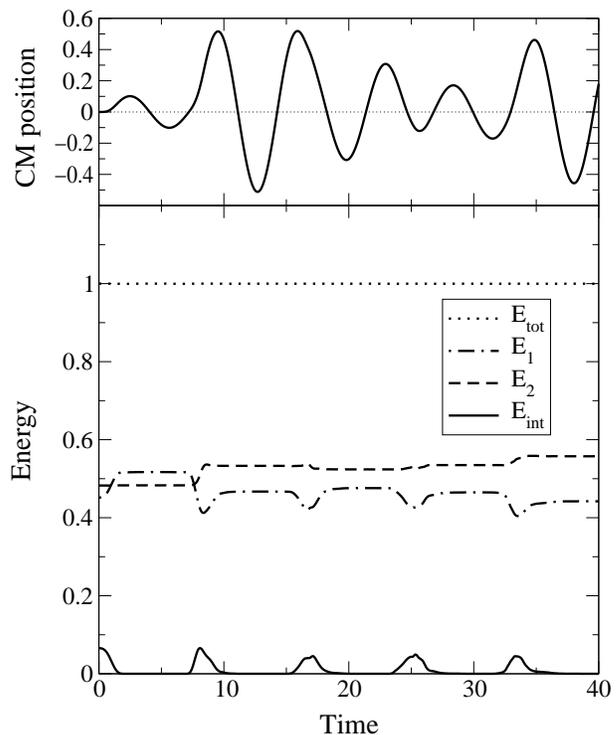}}
\caption{Plot of the center-of-mass oscillation of $|2\rangle$ condensate
(top) and of the various energy terms contributing to 
the total energy of the system (bottom), vs. time. The total energy
(dotted line) is divided as follows: energy of the $|2\rangle$ condensate
(kinetic + potential + mean-field, dashed), energy of the $|1\rangle$ condensate
(dot-dashed), and interaction energy between the two condensates (solid line).
}
\label{fig:energy}
\end{figure}

\begin{figure}
\centerline{\includegraphics[width=8cm,clip=]{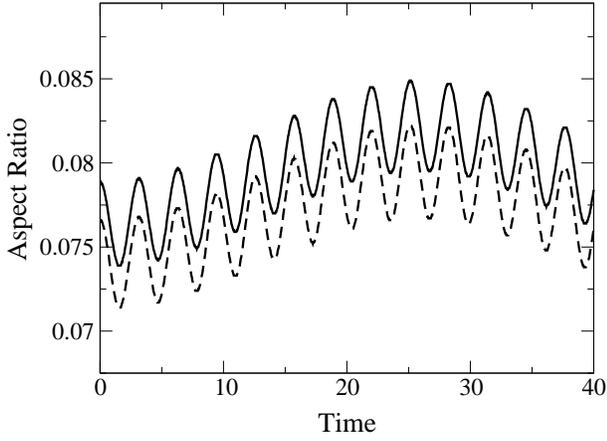}}
\caption{Evolution of aspect ratio for the $|2\rangle$
condensate in the trap, as a function of time (in units of 
$\omega_{\perp2}^{-1}$) when the interaction with the $|1\rangle$
condensate is neglected. The solid line is the GP prediction for 
$[\langle y^2 \rangle / \langle z^2 \rangle ]^{1/2}$, while
the dashed line is $R_{\perp}/Z$ in the Thomas-Fermi limit.  
}
\label{fig:aspect-nonint}
\end{figure}

\begin{figure}
\centerline{\includegraphics[width=8cm,clip=]{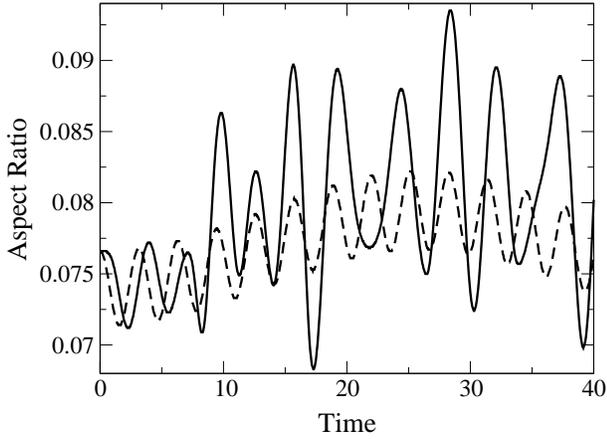}}
\caption{Evolution of aspect ratio for the $|2\rangle$
 condensate in the trap, as a function of time (in units of 
$\omega_{\perp2}^{-1}$). The periodic interaction with the 
$|1\rangle$ condensate is included in the case of the solid 
line and neglected for the dashed line. }
\label{fig:aspect-int}
\end{figure}

\begin{figure}
\centerline{\includegraphics[width=8cm,clip=]{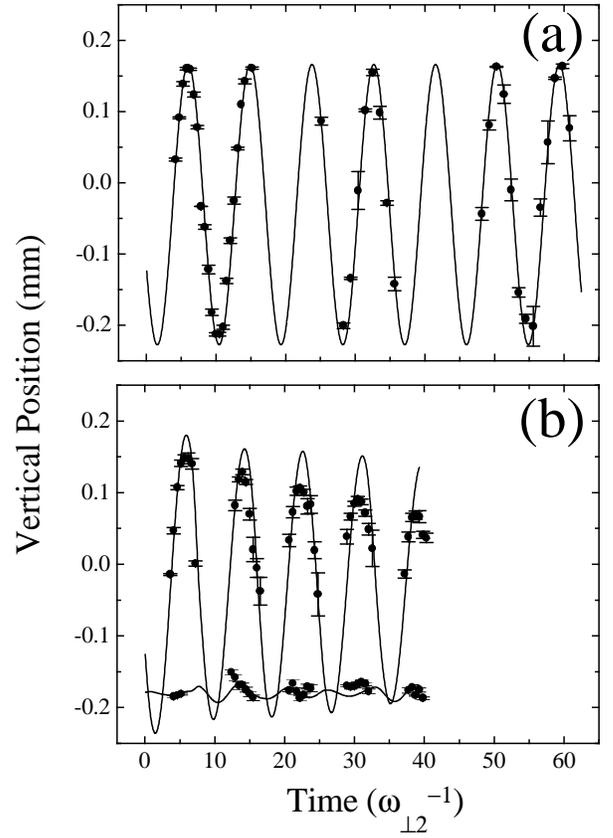}}
\caption{Center-of-mass oscillations as a function of the trapped 
evolution time (in units of $\omega_{\perp2}^{-1}$),  after 
$t_{exp}= 29.3$~ms of ballistic expansion: a) CM oscillations of 
the $|1\rangle$ condensate alone;  b) CM oscillations of both 
condensates, subject to mutual interaction. Theory (solid lines)
is compared with experiments (points). 
}
\label{fig:cm}
\end{figure}
 
\begin{figure}
\centerline{\includegraphics[width=8cm,clip=,]{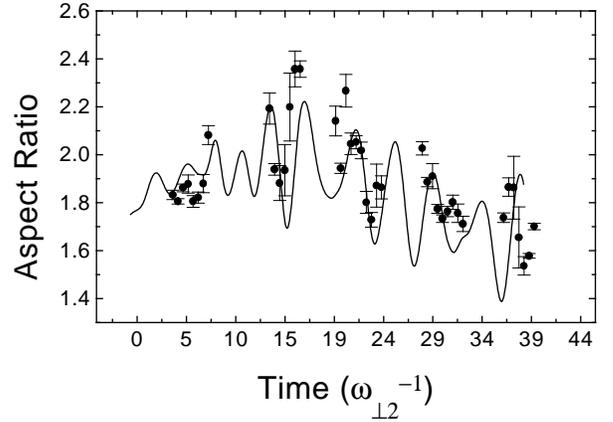}}
\caption{Aspect ratio for the $|2\rangle$ condensate, 
as a function of the trapped evolution time (in units of
$\omega_{\perp2}^{-1}$), 
after $t_{exp}= 29.3$~ms of ballistic expansion.
The experimental points are
compared with the theoretical predictions for the case of two interacting 
condensates.
}
\label{fig:aspect2}
\end{figure}


\begin{references}
\bibitem{note} This shift, which arises from the fact that the 
TF approximation does not include the outer tail of the density 
at the classical turning point, has to be considered in comparing  
theory and experiment, since most of the 
experimental data, as the ones in \protect\cite{maddaloni}, are
extracted by fitting the density distribution with a TF parabola.
For this reason, in Fig.~\protect\ref{fig:aspect2} we 
renormalize the initial aspect ratio given by the GP equation to
the TF value for the same condensate. This normalization, however, 
does not affect the main conclusions of our analysis.
\end{references}
\end{document}